\documentclass[aps,prl,twocolumn,amsmath,amssymb,showpacs]{revtex4-1}
\usepackage{graphicx}% Include figure files
\usepackage{dcolumn}% Align table columns on decimal point
\usepackage{bm}% bold math

\begin{document}

\title{
Spatial solitons in thermo-optical media from the nonlinear \\
 Schr\"odinger-Poisson equation and dark matter analogues
}
\author {Alvaro Navarrete$^{1}$, Angel Paredes$^{2}$,
Jos\'e R. Salgueiro$^{2}$ and
 Humberto Michinel$^{2}$}

\affiliation{
 $^{1}$Departamento de Teor\'\i a do Sinal e Comunicaci\'ons,
Universidade de Vigo, Campus Lagoas-Marcosende, Vigo, ES-36310 Spain.\\
$^{2}$\'Area de \'Optica, Departamento de F\'\i sica Aplicada,
Universidade de Vigo, As Lagoas s/n, Ourense, ES-32004 Spain.}

\begin{abstract}
We analyze theoretically the Schr\"odinger-Poisson equation in two transverse dimensions
in the presence of a Kerr term. The model describes the nonlinear propagation
of optical beams in thermo-optical media and can be regarded as an analogue system for a
self-gravitating self-interacting wave. 
We compute numerically the family of radially symmetric ground state bright stationary solutions
for focusing and defocusing local nonlinearity, keeping in both cases a focusing nonlocal nonlinearity. 
We also analyze excited states and oscillations induced by  fixing the temperature at
the borders of the material.
We provide simulations of soliton interactions,
drawing analogies with the dynamics of galactic cores in the scalar field dark matter scenario.

\end{abstract}


\maketitle

\section{I. Introduction}

Optical  solitons have been a subject of intense research during the last 
decades \cite{Kivshar2003xv,1464-4266-7-5-R02,0034-4885-75-8-086401}.
The interplay of dispersion, diffraction and different types of nonlinearities gives rise to an amazing
variety of phenomena. This 
has  led to an ever increasing control on light propagation and 
to qualitative and quantitative connections to other areas of physics.

The Schr\"odinger-Poisson equation, sometimes also called Schr\"odinger-Newton or 
Gross-Pitaevskii-Newton equation was initially introduced to describe self-gravitating
scalar particles, as a non-relativistic approximation to boson stars \cite{PhysRev.187.1767}.
Since then, it has found application in very disparate physical contexts. 
For instance, it has been used in foundations of quantum mechanics to 
model  wavefunction collapse \cite{diosi,penrose1,penrose2,1367-2630-16-11-115007}, 
in particular situations of
cold boson condensates with long-range interactions \cite{atom1}
or for fermion gases in magnetic fields \cite{PhysRevLett.115.023901}.
In cosmology, it plays a crucial role for two different dark matter scenarios,
namely those of quantum chromodynamic axions \cite{guth} and scalar field dark matter
\cite{1475-7516-2007-06-025,Matos,Marsh:2015xka,schive,schiveprl}
(usually abbreviated as $\psi$DM or SFDM, it also goes under the name of
fuzzy dark matter FDM \cite{witten}).
In nonlinear optics, it can describe the propagation of light in liquid nematic crystals
\cite{PhysRevLett.91.073901,2040-8986-18-5-054006} or thermo-optical media \cite{segev}.

This broad applicability underscores the interest of theoretical analysis of different versions
of the Schr\"odinger-Poisson equation. Moreover, it paves the way for the design of 
laboratory analogues of gravitational phenomena \cite{atom1,segev}. 
It is worth remarking that nonlinear optical analogues have been useful in the past to make
progress in different disciplines, {\it e.g.} the generation of solitons in condensed cold atoms
\cite{PhysRevA.57.3837,StreckerSolitons,Khaykovich1290} or the understanding 
of rogue waves in the ocean \cite{OpticalRogue}. Gravity-optics analogies have
been studied for Newtonian gravity \cite{segev}, aspects of general relativity \cite{segev,Philbin1367}
and even issues related to quantum gravity \cite{Braidotti:2016ido}. In the present context, 
we envisage the
possibility of mimicking certain qualitative aspects of gravitating galactic dark matter waves
in the $\psi$DM model  by studying laser beams in 
thermo-optical media. Although this can only be a partial analogy (the optical dynamics is 1+2 dimensional
instead of 1+3), it is certainly appealing and can lead to novel views on both sides.

Consequently, this article focuses on the following system of equations:
\begin{eqnarray}
i\frac{\partial \psi}{\partial z}&=&-\frac12 \nabla^2 \psi - \lambda_K |\psi|^2\psi +
\Phi \psi 
\label{parax2}\\
\nabla^2 \Phi &=& C |\psi|^2
\label{poisson2}
\end{eqnarray}
The constant $C$ can be fixed to any value and we will choose $C=2\pi$.
$|\psi|^2$ represents the laser beam intensity and $\Phi$ corresponds to
the temperature, see section II for their precise definitions. In the
$\psi$DM escenario, $|\psi|^2$ is associated to the dark matter density and
$\Phi$ to the gravitational potential.

All coefficients in (\ref{parax2}), (\ref{poisson2}) have
been rescaled to bring the expressions to their canonical form and all quantities are dimensionless.
This rescaling can be performed without loss of generality, see
section II and the appendix for the details of
the relation to dimensionful parameters. 
Notice that the wavefunction $\psi$ is complex while the potential $\Phi$ is real.
The  coordinate $z$ is the propagation distance in optics and plays the role of
time in condensed matter waves. The Laplacian acts on two transverse dimensions $d=2$.
We constrain ourselves to the case in which the Poissonian interaction is attractive and
thus fix a positive sign for the $\Phi \psi$ term.
The constant $\lambda_K=\pm 1$ is related to the focusing (defocusing) Kerr nonlinearity
for positive (negative) sign. For matter waves, it is proportional to the s-wave scattering
length which leads to attractive (repulsive) local interactions. We remark that in the 
$\psi$DM model of cosmology, this term is sometimes absent \cite{schive}
 but there are also numerous
works considering $\lambda_K \neq 0$, {\it e.g.} \cite{PhysRevD.53.2236,Goodman,guzman1}.

Equation (\ref{poisson2}) has to be supplemented with boundary conditions. 
This is a compelling property of $d=2$ since it allows to partially control the dynamics by
tuning the boundary conditions, as  demonstrated in 
\cite{PhysRevLett.95.213904,Alberucci:07,Alberucci:07a,Alfassi:07}.
In this aspect, there is a marked difference with the 
three-dimensional case $d=3$, in which
 $\Phi \to 0$ at spatial
infinity if the energy distribution $|\psi|^2$ is confined to a finite region.

Another point that raises interest on the  study of 
equations (\ref{parax2}) and (\ref{poisson2}) is that they
include competing nonlinearities, one of which is
nonlocal. 
Different kinds of competing nonlinearities have been thoroughly studied since they 
can improve the tunability of optical media and lead to rich dynamics, see {\it e.g}
\cite{PhysRevLett.102.203903,Gurgov:09,
Laudyn:15,PhysRevA.83.053838,Cavitation,PhysRevLett.116.163902,PhysRevLett.115.253902, 
0295-5075-98-4-44003}.
On the other hand, it
 is well known that nonlocal interactions can stabilize nonlinear solitary waves since
they tend to arrest collapse
\cite{PhysRevE.62.4300,PhysRevE.66.046619,1464-4266-6-5-017}.
Moreover, nonlocality can lead to long-range interaction between solitons \cite{segev2}
and other interesting phenomena as, for instance the stabilization of
multipole solitons \cite{Kartashov:06,Lopez-Aguayo:06}.

The existence of robust solitons for the Schr\"odinger-Poisson equation with
$d=3$ has been
demonstrated and their properties have been thoroughly analyzed 
\cite{diosi,penrose,harrison,atom1,atom2,PhysRevA.78.013615,PhysRevA.82.023611,0004-637X-645-2-814,Chavanis,Pop}.
However, the
 two dimensional case  has received less attention, although we must
stress that  relevant results in similar contexts without the Kerr term can be found in 
\cite{harrison,PhysRevE.71.065603,PhysRevA.76.053833,PhysRevA.91.013841,Alberucci:14,breather}.
Here, we intend to close this gap by performing a systematic analysis of the basic stationary
solutions of (\ref{parax2}), (\ref{poisson2}) and by simulating some simple interactions
between them.

In section II, we discuss the implementation of equations (\ref{parax2}), (\ref{poisson2}) in
nonlinear optical setups. Section III is devoted to the analysis of the 
simplest
eigenstates of
these equations, namely the spatial optical solitons.
The cases of focusing and defocusing Kerr nonlinearities are discussed in turn.
 In section IV, we comment on
their interactions and on analogies with dark matter theories. 
The different sections are (mostly) independent and can be read separately.
Section V summarizes our
findings.

\section{II. The optical setup}

The Schr\"odinger-Poisson system, without the Kerr term, describes the propagation
of a continuous wave laser beam in a thermo-optical medium, see {\it e.g.} \cite{segev}
and references therein. In this section,
we briefly review the formalism in order to make the discussion reasonably self-contained
and to fix notation.
Then, we argue that the
Kerr term can play a significant role in certain situations.
Finally, we discuss some details of interest for a possible experimental implementation
and the conserved quantities.

\subsection{a. Formalism}

The paraxial propagation equation for a beam 
of angular frequency $\omega$
in a medium with 
refractive index $n=n_0 + \Delta n$ is given by:
\begin{equation}
-2ik_0 n_0\frac{\partial A}{\partial \tilde z}= \tilde \nabla^2 A + 2 \Delta n k_0^2 n_0 A
\label{parax1}
\end{equation}
where we have assumed that $n_0$ is a constant and  
neglected terms of order ${\cal O}\left( \Delta n^2\right)$.
We denote with a tilde the dimensionful coordinates such that
the Laplacian is
$\tilde \nabla^2 \equiv \partial_{\tilde x}^2+ \partial_{\tilde y}^2$.
The electric field is $E={\rm Re}[A e^{i( n_0 k_0 \hat z-\omega \tilde t)}]$,
the intensity is given by $I=\frac{n_0}{2\eta_0}|A|^2$ where
$\eta_0=\sqrt\frac{\mu_0}{\epsilon_0} $ and $k_0=\omega/c$ is the wavenumber in
vacuum. In the paraxial approximation, the electromagnetic wave envelope $A$ 
is assumed to vary slowly at the scale of the wavelength $\partial_{z}^2 A \ll k \partial_z A$.
We will consider a model in which $\Delta n$ is the sum of an optical Kerr term
$\Delta n_{K} = n_2 I=\frac{n_2n_0}{2\eta_0}|A|^2$
and of a thermo-optical variation of the refractive index
$\Delta n_{T} = \beta \Delta T$ where $\beta$ is the thermo-optic coefficient
which we assume to be positive
and the temperature is defined as $T=T_0 + \Delta T$ where $T_0$ is a fiducial
constant.

We now write down the equation determining the temperature distribution in the 
material. In a stationary situation, it is given by
$\kappa \tilde \nabla^2 T = q$,
where $\kappa$ is thermal conductivity (with units of $\frac{W}{m\,K}$)
and $q$ the heat-flux density of the source (power exchanged per unit volume),
which comes from the absorption of the beam in the material $q=-\alpha\,I$
where $\alpha$ is the linear absorption coefficient of the optical medium.
Thus,
\begin{equation}
{\kappa} \tilde \nabla^2 \Delta T = -\frac{\alpha n_0}{2 \eta_0} |A|^2
\label{poisson1}
\end{equation}
We are taking here a two dimensional Laplacian, therefore assuming
$\frac{\partial^2 \Delta T}{\partial \tilde z^2} \ll  \tilde \nabla^2 \Delta T $.
This is a kind of paraxial approximation for the temperature distribution
motivated by the paraxial distribution of the source beam.
Notice, however, that the neglected term
may play a role in non-stationary situations under certain circumstances
\cite{PhysRevA.91.013841}.

We can rewrite (\ref{parax1}), (\ref{poisson1}) in dimensionless form
(\ref{parax2}), (\ref{poisson2}) 
taking $\lambda_K$ to be the sign of $n_2$, $C=2\pi$ and performing the
following rescaling (see the appendix):
\begin{eqnarray}
&&\tilde z=\frac{2\pi\kappa n_0 |n_2| k_0}{\alpha \beta} z, \qquad
(\tilde x,\tilde y)=\sqrt\frac{2\pi\kappa |n_2|} {\alpha \beta}(x,y)\ \nonumber \\
&&A=\sqrt\frac{ \eta_0 \alpha \beta}{\pi \kappa n_0^2 |n_2|^2 k_0^2}  \psi, \quad
\Delta T  = - \frac{\alpha} {2\pi \kappa n_0 |n_2| k_0^2}  \Phi\
\label{change}
\end{eqnarray}
The power of the beam $\tilde P=\int I d\tilde x d\tilde y$ is given by:
\begin{equation}
\tilde P=\frac{P}{n_0 |n_2| k_0^2}  \equiv \frac{1}{n_0 |n_2| k_0^2} \int |\psi|^2 dx dy
\label{power}
\end{equation}
The limit $P \ll 1$ corresponds to
negligible Kerr nonlinearity, a fact that  will be made explicit in section III when
discussing the eigenstates.

\subsection{b. The Kerr term}

The goal of this paper is to perform a general analysis of the
dimensionless equations (\ref{parax2}), (\ref{poisson2}), which can be associated to
a particular physical scenario through Eqs. (\ref{change}), (\ref{power}) or, in general, Eq.
(\ref{changegen}) in the appendix. Nevertheless, it is interesting to consider a particular
case in order to provide benchmark values for the physical quantities.
Thus, let us quote the values associated to the experiments in 
\cite{PhysRevLett.95.213904,segev2,segev}, in which a continuous wave laser beam
with a power $\tilde P$ of a few watts and
 $\lambda=488$nm propagates through lead glass with
$\kappa$=0.7 W/(m \ K), $\beta$=14 $\times 10^{-6}$ K$^{-1}$,
 $n_0=1.8$, $\alpha$=0.01 cm$^{-1}$ (values taken from
\cite{PhysRevLett.95.213904}) and $n_2=2.2\times 10^{-19}$m$^2$/W \cite{Gurgov:09}.
In this setup, $P \approx 10^{-4}$ and the Kerr term is inconsequential.
In order to motivate the inclusion of this term in (\ref{parax2}), it is worth commenting
on different experimental options to increase $P$.

The first possibility is to treat the material in order to enlarge $|n_2|$. This can be done by 
doping it with metallic nanoparticles \cite{singh} and/or ions \cite{Can-Uc:16}. Another option is to use
a pulsed laser. The thermo-optical term, being a slow nonlinearity,
 mostly depends on the average power and therefore does not change  much
 with the temporal structure of the pulse. On the other hand, the Kerr term does of course 
 depend on the peak power. Thus, for a pulsed laser, we can use the same formalism
 (\ref{parax2}), (\ref{poisson2}) and, compared to a continuous wave laser of the same average
 power, it amounts to enhancing $n_2$ by a factor which is approximately $(\tau R_r)^{-1}$
 where $\tau$ is the pulse duration and $R_r$ the repetition rate.
 In fact, this kind of interplay between slow (nonlocal) and fast (local) nonlinearities 
 has been demonstrated for spatiotemporal solitons, also called light bullets
 \cite{PhysRevLett.102.203903,Gurgov:09}.
 In our case,
Eqs. (\ref{parax2}), (\ref{poisson2}) do not include the temporal dispersion and would not be 
valid for very short pulses, but are well suited for, {\it e.g.}, Q-switched lasers, where
both kind of nonlinearities can be comparable for the spatial dynamics of the beam.

\subsection{c. Measurable quantities and boundary conditions}

We now briefly comment on certain  details of interest for an eventual experimental 
implementation. We do so by quoting the techniques employed in experiments of laser
propagation in lead glass, see \cite{PhysRevLett.95.213904,segev} and references therein.

The first question is what observables can actually be measured. As in 
\cite{segev}, we envisage the possibility of taking images of the laser power distribution at the
entrance and exit facets. Below, we present plots of the evolution of the spatial profile of
the beam at different values of the propagation distance $z$. They would correspond to
propagation within sections of the thermo-optical material of different lengths, with the
rest of conditions fixed. 
Notice that,
in the absence of Kerr term ($\lambda_K=0$), there is a
scaling symmetry $\gamma^2  \psi(\gamma x,\gamma y, \gamma^2 z)$,
$\gamma^2   \Phi(\gamma x,\gamma y, \gamma^2 z)$ solves (\ref{parax2}), (\ref{poisson2}) for
any $\gamma$ if it is a solution for $\gamma=1$. 
Thus, in this case,
different adimensional
propagation lengths can be studied just by changing the initial power and width of the beam,
and not the medium itself.

Of course, it would be of interest to measure the intensity profile 
within the material, but we are unaware of techniques that can achieve that goal without
distorting the beam propagation itself. We are also
unaware of techniques to measure the temperature distribution within the material and, thus,
$\Delta T$ can be estimated through the modeling equations but can only be 
indirectly
compared
to measurements.

A second important point is that of boundary conditions for the Poisson field or, in physical terms,
how to fix the temperature at the borders of the material. This can be done by thermally
connecting the borders to heat sinks at fixed temperature 
which exchange energy with the optical material \cite{PhysRevLett.95.213904}. Therefore, 
in a typical experiment, the mathematical problem is supplemented with Dirichlet boundary conditions.
The boundary value of $\Phi$ can be tuned to be different at different positions of the perimeter,
giving rise to a turning knob useful to control the laser dynamics from the exterior of the
sample \cite{PhysRevLett.95.213904}. However, for simplicity, in this work
we will consider $\Phi$ at the
border to be constant.
Non-Dirichlet boundary conditions for $\Phi$ are also feasible in experimental implementations.
For instance, if the edge of the material is thermally isolated, Neumann boundary conditions
are in order.

It is also worth commenting on the behavior of the electromagnetic wave at the borders. If
 at most a negligible fraction of the optical energy reaches the boundary facets 
  that are
parallel to propagation, the boundary conditions for $\psi$ used in  computations
become irrelevant. However, as we show below, capturing some aspects of
dark matter evolution requires long propagations during which, unavoidably, part of the
radiation does reach the border. In an actual experiment, the easiest is to have reflecting boundary
conditions, with the interface acting as a mirror for light. Physically, for a dark matter analogue,
it would be better in turn to avoid reflections and therefore to have open or absorbing
boundary conditions. This can be accomplished by attaching an absorbing element with the
same real part of the refractive index as the bulk material. We will come back to this 
question in section IV.

\subsection{d. Conserved quantities}

We close this section by mentioning the   conserved quantities upon propagation in $z$. 
We assume here that $\psi$ is vanishingly small near the boundary of the sample
and that generic Dirichlet conditions hold for $\Phi$.
It is then straightforward to check 
from (\ref{parax2}), (\ref{poisson2})
that the  norm $N= \int |\psi|^2 d^2{\bf x}$ 
and
hamiltonian:
\begin{equation}
H= \frac12 \int \left(
\vec\nabla \psi^* \cdot \vec\nabla \psi -  \lambda_K |\psi|^4 +
 \Phi |\psi|^2
\right)dxdy
\end{equation}
do not change during evolution in $z$.

\section{III. Radially symmetric bright solitons}

In this section, we study stationary solutions
of Eqs. (\ref{parax2}), (\ref{poisson2})
 of the form $\psi=e^{i\mu z} f(r)$,
$\Phi = \phi(r)$ where we have introduced $r=\sqrt{x^2+y^2}$.
With a usual abuse of language, we refer to these solutions as solitons.
The system gets reduced to:
\begin{eqnarray}
\mu f(r)&=&\frac12 \frac{d^2 f(r)}{dr^2} +\frac{1}{2r}\frac{df(r)}{dr} +\lambda_K f(r)^3 -
 \phi(r) f(r) \nonumber\\
\label{parax4}\\
 2 \pi f(r)^2&=& \frac{d^2 \phi(r)}{dr^2} + \frac{1}{r}\frac{d\phi(r)}{dr} 
\label{poisson4}
\end{eqnarray}
Equation (\ref{poisson4}) has to be supplemented
with a boundary condition for $\phi(r)$ since, unlike the $d=3$ case, it is not possible
to require that $\lim_{r\to \infty} \phi(r) = 0$. 
We consider a
boundary condition that preserves radial symmetry.
Non-radially symmetric boundary conditions lead to non-radially symmetric
solitons \cite{PhysRevLett.95.213904}, whose systematic study we leave for future work.
Therefore, we impose
$\phi(R)=\phi_R$, where $\phi_R$ is an arbitrary constant and
 $R$ is much larger than the bright soliton radius
$R \gg r_{sol}$. 
The particular values of $R$ and $\phi_R$ are unimportant because
for $r\gg r_{sol}$, the optical field vanishes $f(r) \approx 0$
and the potential reads $\phi = \phi_R + P \, \log(r/R)$, where
$P$ is the adimensional power defined in eq. (\ref{power}). Thus, changing
$\phi_R$ and $R$ only amounts to adding a constant to $\phi$ which can be absorbed
as a shift in $\mu$, while the beam profile $f(r)$ is unaffected.
However, in order to compare the propagation constant $\mu$ of different
solutions, it is important to compute them with the same convention.
In our computations, we take, without loss of generality $\phi_R=0$, $R=100$.
For large $r$, the function $f(r)$ decays as
$ \exp (-r\sqrt{2 P\log (r/R)} ) $. We remark that in the case of defocusing nonlocal nonlinearity
there are no decaying solutions of this kind and, accordingly, there are no bright solitons.

Enforcing regularity at $r=0$, we find the following expansion, in terms of
two constants $f_0$ and $\varphi_0$. The propagation constant $\mu$ can be
absorbed as a shift in $\phi$ for the computation, taking $\varphi(r)=\phi(r)+\mu$:
\begin{eqnarray}
f(r)&=& f_0 + \frac{f_0}{2}(\varphi_0 - \lambda_K f_0^2) r^2 + {\cal O}(r^4)
\nonumber\\
\varphi(r)&=& \varphi_0 + \frac{\pi}{2}f_0^2 r^2  + {\cal O}(r^4)
\end{eqnarray}

\subsection{a. Focusing Kerr term}

We start discussing the Kerr focusing case $\lambda_K=1$.
For any positive value of $f_0$ there is a discrete set of values
$\varphi_{0,i}(f_0)$ which yield
normalizable solutions with $i=0,1,2,\dots$ nodes for $f(r)$. 
These values can be found numerically, for instance using a simple shooting technique.
For each solution, the value of $\mu$ is computed from the boundary condition for
$\phi(R)$. 

\begin{figure}[h!]
\begin{center}
\includegraphics[width=.9\columnwidth]{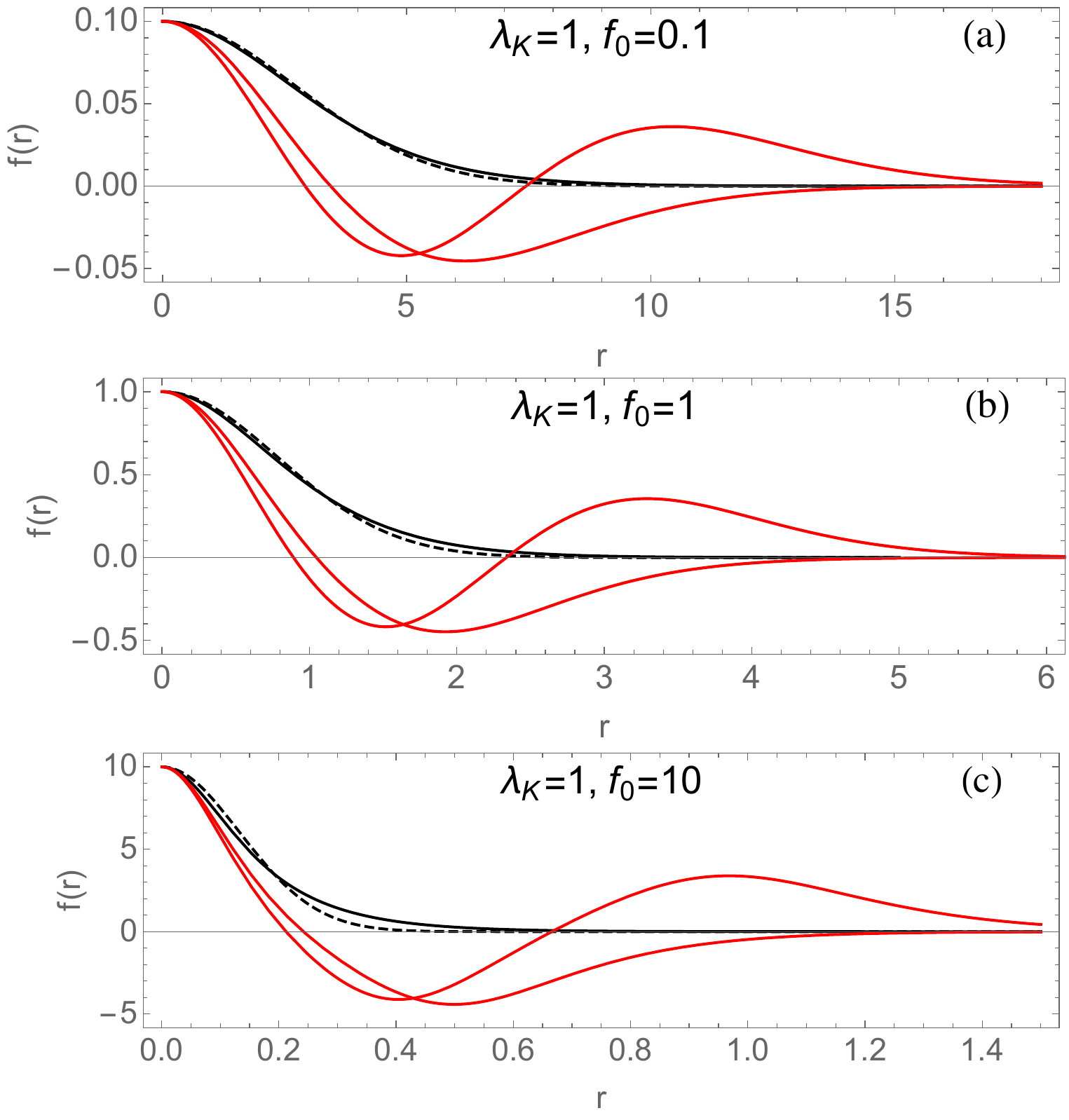}
\end{center}
\caption{[Color online] The ground state solution for $\lambda_K=1$, 
$f_0=0.1,1,10$ compared to a gaussian (dashed 
line). In red, the solutions with one and two nodes in each case.}
\label{fig1}
\end{figure}

In figure \ref{fig1}, we plot the $f(r)$ profiles of the solutions with $i=0,1,2$
for different values of $f_0$ and $\lambda_K=1$. 
We compare the ground state solutions $i=0$
to gaussians with the same value of $f(r=0)$ and norm. Gaussians are a usual
approximate
trial function for soliton profiles in nonlocal media \cite{Snyder1538} and the figure shows that
in the present case, the approximation is rather precise and becomes better for smaller $f_0$.

\subsubsection{Ground state}

\begin{figure}[h!]
\begin{center}
\includegraphics[width=\columnwidth]{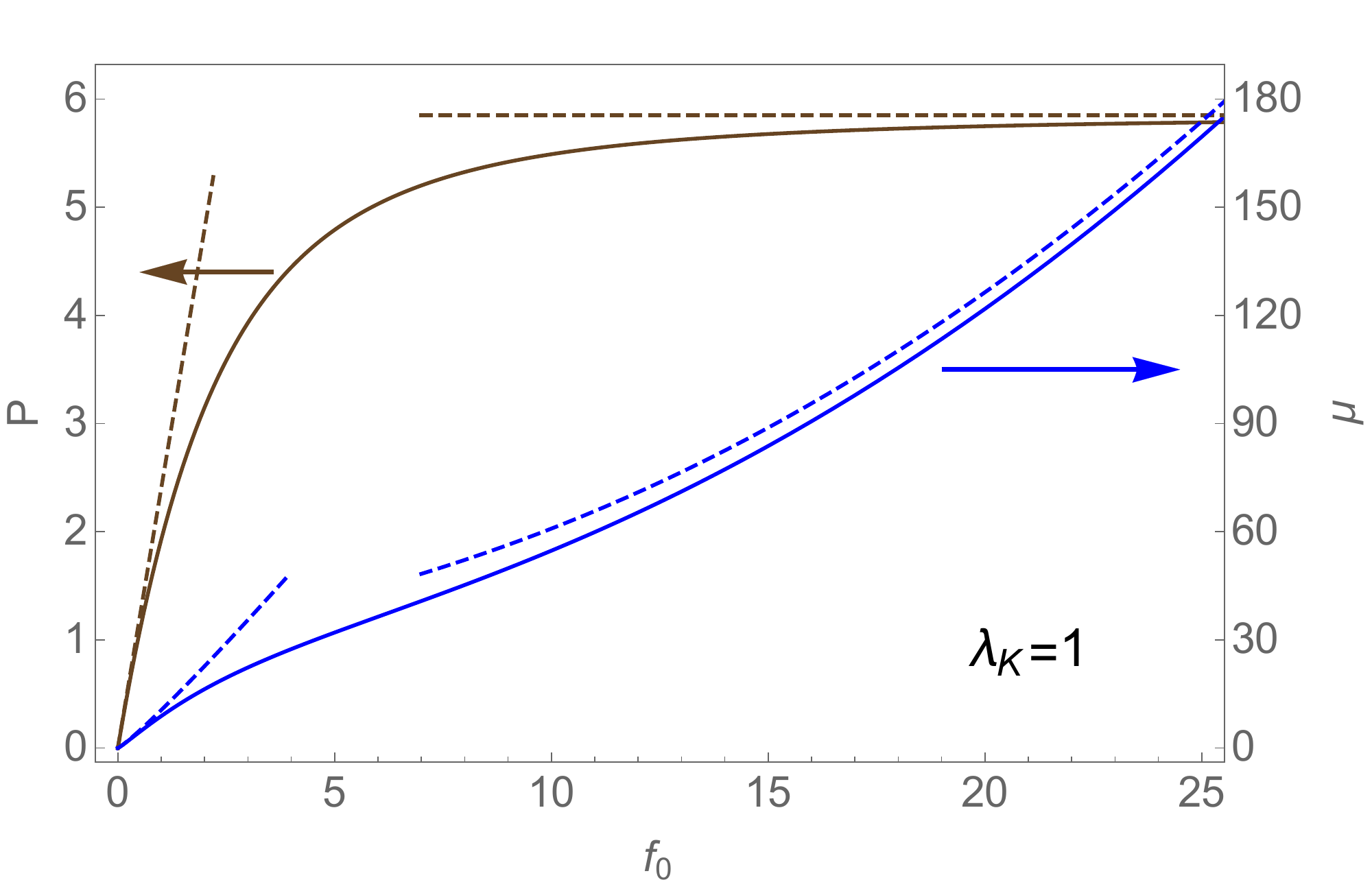}
\end{center}
\caption{[Color online] The adimensional power $P$ and propagation constant $\mu$ as a function of $f_0$ 
for the Schr\"odinger-Poisson equation with focusing Kerr term $\lambda_K=1$. Dashed lines
represent the asymptotic behaviors described in the text.}
\label{fig2}
\end{figure}

In figure \ref{fig2}, we depict how the power and propagation constant vary
within the family of ground state solutions with $\lambda_K=1$, that
interpolate between the solution without Kerr term for $f_0 \to 0$  
and the one with only Kerr term, namely the
Townes profile \cite{PhysRevLett.13.479}, for $f_0 \to \infty$.
Explicitly, for small $f_0$, we have $P\approx 2.40 f_0$, $\mu\approx 10.53 f_0 + 1.20 f_0 \log f_0$,
where the logarithmic term is related to the boundary condition for $\phi(R)$.
For the large $f_0$, we have $P\approx 5.85 $ and $\mu \approx 0.205 f_0^2 + 
\left(26.9 +5.85 \log f_0 \right)$.
The $\mu$ is the sum of the one of the Townes solution plus a term
coming from the value of $\phi(r)$ at small $r$. Notice that
the light intensity is confined to a small region in $r < r_{sol}$ where $\phi$ reaches its minimum.
Away from it, $\phi(R) \approx \phi(r_{sol}) + P \log\frac{R}{r_{sol}}$. Thus, from 
our boundary condition
$\phi(100)=0$,
we find $\phi(r_{sol})=-P \log\frac{100}{r_{sol}}$, where $r_{sol} \approx f_0^{-1}$ for
the Townes profile.

Both $P(f_0)$ and $\mu(f_0)$ are monotonically increasing functions.
Thus $\frac{dP}{d\mu}$ is always positive within the family and all solutions are stable
according to the Vakhitov-Kolokolov criterion. Clearly this derivative approaches zero
for large $f_0$, as expected for the Townes profile.

\subsubsection{Excited states and details on evolution algorithms}

\begin{figure}[h!]
\begin{center}
\includegraphics[width=.9\columnwidth]{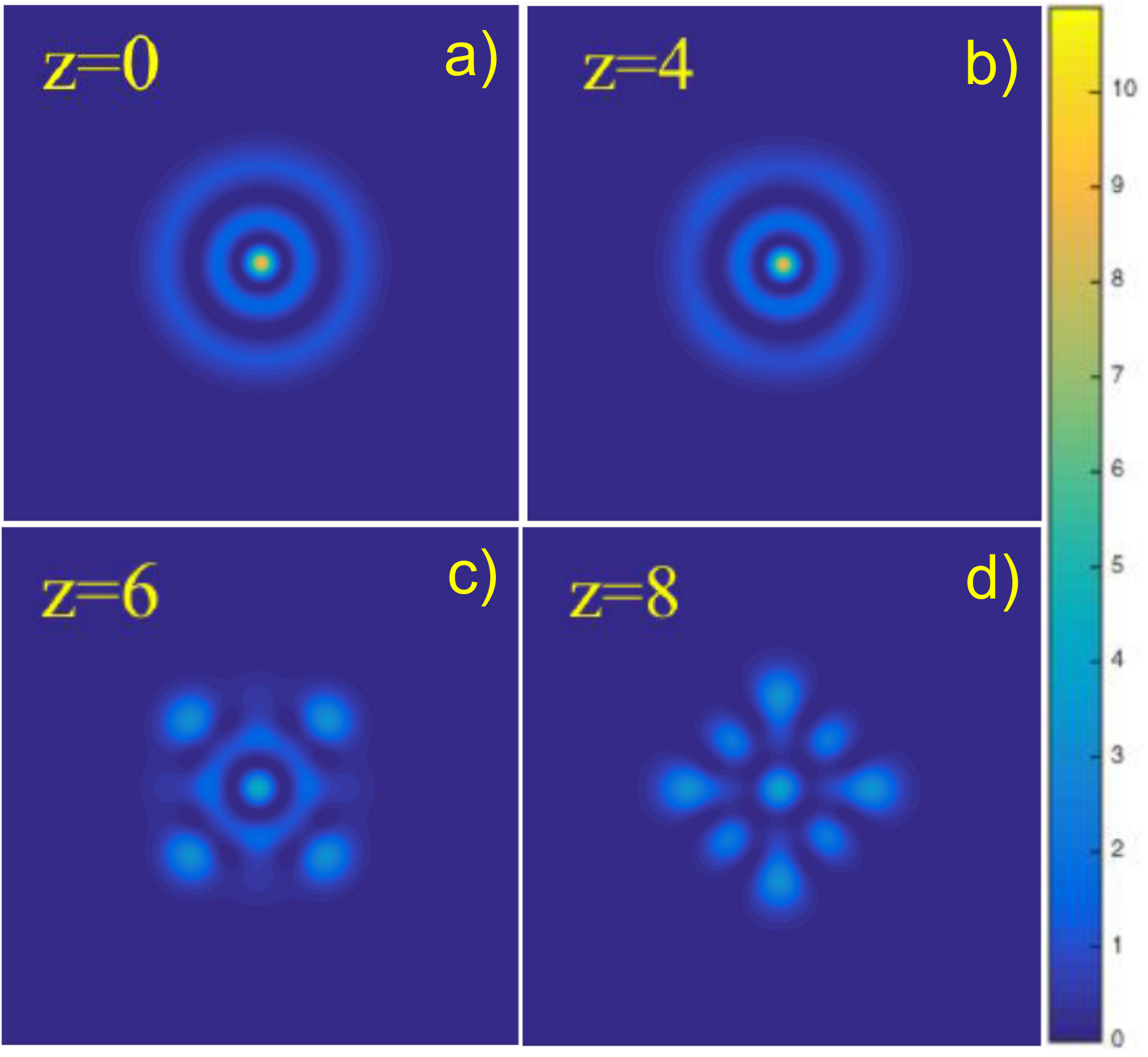}
\end{center}
\caption{[Color online] An illustration of the
initial stages of the
 evolution of an unstable solution with two nodes, $\lambda_K=1$
and $f_0=3$. The size of the images is 10 $\times$ 10 (dimensionless units). 
Boundary conditions
$\phi=0$ are set in the sides of a 20 $\times$ 20 square.}
\label{fig3}
\end{figure}

Let us now turn to excited states $i \geq 1$. In figure \ref{fig3}, we show
an example of the disintegration of the solution with $f_0=3$ with two nodes.
The computation of figure \ref{fig3} and the rest of dynamical simulations 
 displayed in this paper are
performed setting boundary conditions in the sides of a square.
The reason is that radial symmetry is not usually preserved by the actual pieces of
 thermo-optical material used in experiments \cite{segev,PhysRevLett.95.213904}.
 In order to compare to the dark matter scenario, the best would be to fix the
 boundary conditions to match the monopolar contribution of the Poisson field
 sourced by the energy distribution. In a non-cylindrical sample material, this requires
 generating a space dependent particular temperature distribution at the boundary, 
 which seems difficult to implement experimentally. Thus, we still fix constant
 $\Phi$ conditions at the borders of the square.
 We remark that radial symmetry is broken
in a soft way if the square is much larger than the beam size and the main
features of the evolution are not affected by this mismatch. 
Our simulations are performed using the beam propagation method 
to solve Eq. (\ref{parax2}) and  a finite difference scheme to
solve Eq. (\ref{poisson2}) at each step. Convergence of the method has been checked
by comparing simulations with different spacing for the spatial  computational 
grids and steps in $z$.

We have not found any excited state stationary solution that preserves its shape
for long propagation distances.
Fig. \ref{fig3} shows the initial stages of the disintegration of an unstable solution.
Following the propagation to larger $z$, the system typically tends towards a ground state
solution, surrounded by some radiation that takes the excess energy. This is similar
to what we will discuss in section IV.c. The analogue behavior in three dimensions was
analyzed in \cite{PRD69}.
When the total power is above the Townes critical value $P\approx 5.85$, 
 the evolution can eventually
result  in collapse, with $\psi$ diverging at finite $z$. When the beam profile becomes extremely
narrow, the nonlocal term is negligible and the collapse is equivalent to that with only
focusing Kerr term.

\subsubsection{Oscillations around the center of the material}
 
The radially symmetric solutions we have discussed require that the center of the thermo-optical
material coincides with the center of the soliton. 
In a first approximation,
if the light beam is 
shifted from the center of the 
material, the beam profile remains unchanged but its center feels a refractive index gradient which
induces an oscillation \cite{Alfassi:07,Alberucci:07,Alberucci:07a,Shou:09}. 
One can think of this phenomenon as a self-force mediated by boundary
conditions. It can be understood in terms in terms of Green's function for two dimensional 
Laplace equation on the disk.
\begin{equation}
G (\rho,\theta)= \frac{P}{2} \log \frac{\hat\rho^2 + \rho^2 - 2 \hat\rho \rho \cos(\theta - \hat\theta)}
{R^2 + \hat\rho^2 \rho^2/R^2 - 2 \hat\rho \rho \cos(\theta - \hat\theta)}
\label{GreenFunc}
\end{equation}
This expression solves (\ref{poisson2}) for a point source of power $P$
placed at $(\hat x,\hat y)$, namely
$|\psi|^2 =P  \delta(x-\hat x) \delta(y-\hat y)$
 and satisfies the boundary condition $\phi(R)=0$. We have introduced
$\rho^2 = x^2 + y^2\leq R^2$, $\hat \rho^2 = \hat x^2+ \hat y^2 < R^2$,  $\theta= \arctan (y/x)$,
and $\hat \theta = \arctan(\hat y/ \hat x)$.
The Green function (\ref{GreenFunc}) is computed
by considering an image at a distance $R^2 / \hat \rho$ from the center of the disk.
We can think of the self-force mediated by boundary conditions as the force performed by 
the image on the source \cite{Shou:09}.

Without loss of generality, consider a soliton
with center at  $x=x_s$ with $y_s=0$.
Taking the gradient of the potential 
generated by the image, keeping only the leading terms  in $|x_s| / R < 1$
and using Ehrenfest theorem we find the approximate expression: 
\begin{equation}
\frac{d^2 x_s}{dz^2} \approx - 2P\frac{x_s}{R^2} - 2P \frac{x_s^3}{R^4}
\label{d2xx}
\end{equation}
where $x_s$ is the position of the soliton at propagation distance $z$.
Notice that this restoring force induced by boundary conditions becomes negligible for
large $R$ and fixed $x_s$. This means that, if the electromagnetic wave is confined in a 
given region, the role of boundaries diminishes when the piece of optical material is
taken wider.

From (\ref{d2xx}), it is immediate to infer a periodic motion with a
period in the propagation distance $Z\approx  \left(\frac{\sqrt2 \pi R}{\sqrt{P}} - \frac{3\pi x_i^2}
{4\sqrt{2P} R} \right)$ for a soliton initially at rest at $x=x_i$.

We have performed a series of simulations, with boundary
conditions $\phi = 0$ set at the boundary of a square of side $L$,
by placing solitons of different powers initially displaced a distance $x_i$ from the center of the square.
As expected from the discussion above, there is an oscillation induced by the boundary
conditions. The period of the oscillation follows the same kind of dependence on
$P$, $L$ and $x_i$ as the one found analytically for the disk.
From our numerics, we infer that the period in $z$ of the oscillation is:
\begin{equation}
Z\approx \left(3.38\frac{L}{\sqrt{P}} - 8.2\frac{ x_i^2}
{\sqrt{P} L} \right)
\label{Zmodel}
\end{equation}

\begin{figure}[h!]
\begin{center}
\includegraphics[width=\columnwidth]{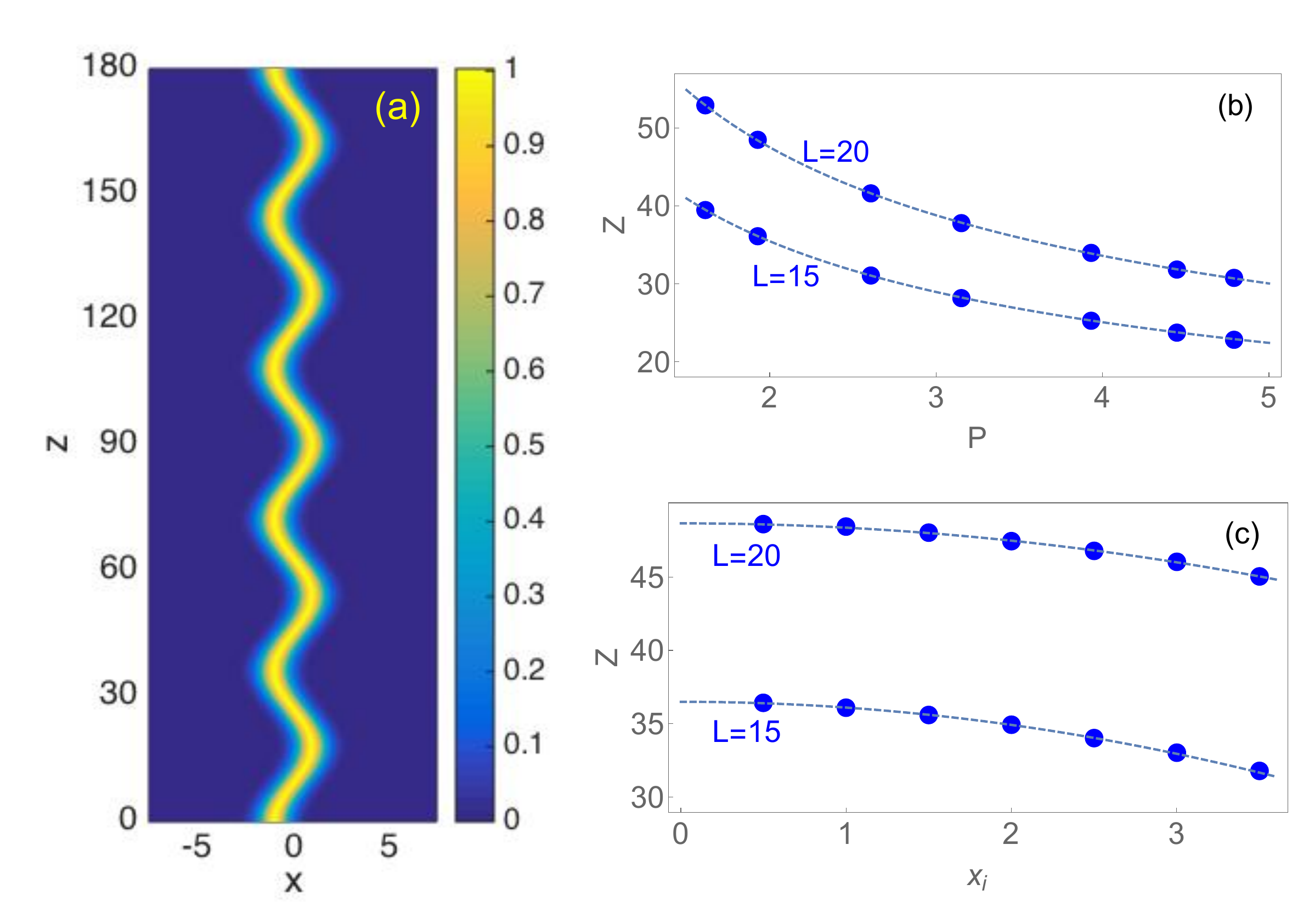}
\end{center}
\caption{[Color online]  Panel a) shows the oscillation in the $x$-direction for a particular example
$f_0=1$ ($P\approx 1.93$), 
$x_i=-1$, $L=15$. We represent a contour plot $|\psi|^2 (x,y=0,z)$. Panel b) represents
examples of 
the numerically computed oscillation period with $x_i=-1$ for two values of $L$, compared to the
model (\ref{Zmodel}) (dashed line). Panel c) is a similar comparison as a function of $x_i$ with
$P=1.93$ fixed.
}
\label{fig4}
\end{figure}

In figure \ref{fig4} we depict an example of the oscillation found by numerically solving 
the evolution equations and a comparison of Eq. (\ref{Zmodel}) with the computed value
of $Z$ for several cases.

\subsection{b. Defocusing Kerr term}

\begin{figure}[h!]
\begin{center}
\includegraphics[width=.9\columnwidth]{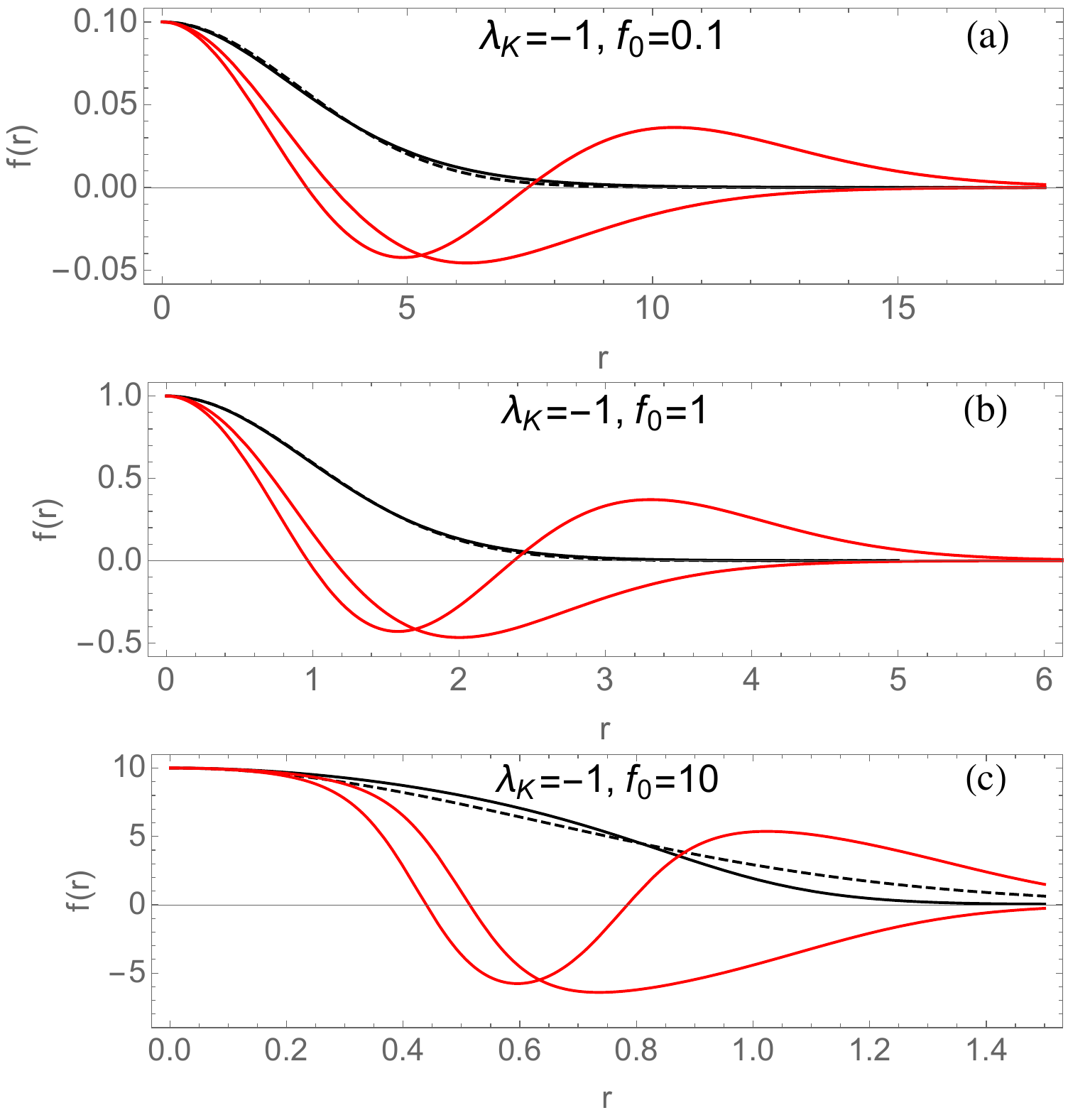}
\end{center}
\caption{[Color online] The ground state solution for $\lambda_K=-1$, 
$f_0=0.1,1,10$ compared to a gaussian (dashed 
line). In red, the solutions with one and two nodes in each case.}
\label{fig5}
\end{figure}

We now turn to the  case of defocusing Kerr nonlinearity $\lambda_K=-1$.
As in the previous case, for any  $f_0>0$ there is a discrete set of values
$\varphi_{0,i}(f_0)$ which yield
normalizable solutions with $i=0,1,2,\dots$ nodes for $f(r)$. 
Again, the ground state is always stable while we have not found stable excited solutions.
Figure \ref{fig5} shows several examples. 
Unsurprisingly, the solutions with $f_0=0.1$ are almost indistinguishable from those
in figure \ref{fig1}, since the $f_0 \to 0$ limit corresponds to negligible Kerr term.
On the other hand, for large $f_0$, the difference becomes apparent. Curiously,
in the intermediate case with $f_0=1$, the numerical solution is really similar to a gaussian.

\begin{figure}[h!]
\begin{center}
\includegraphics[width=\columnwidth]{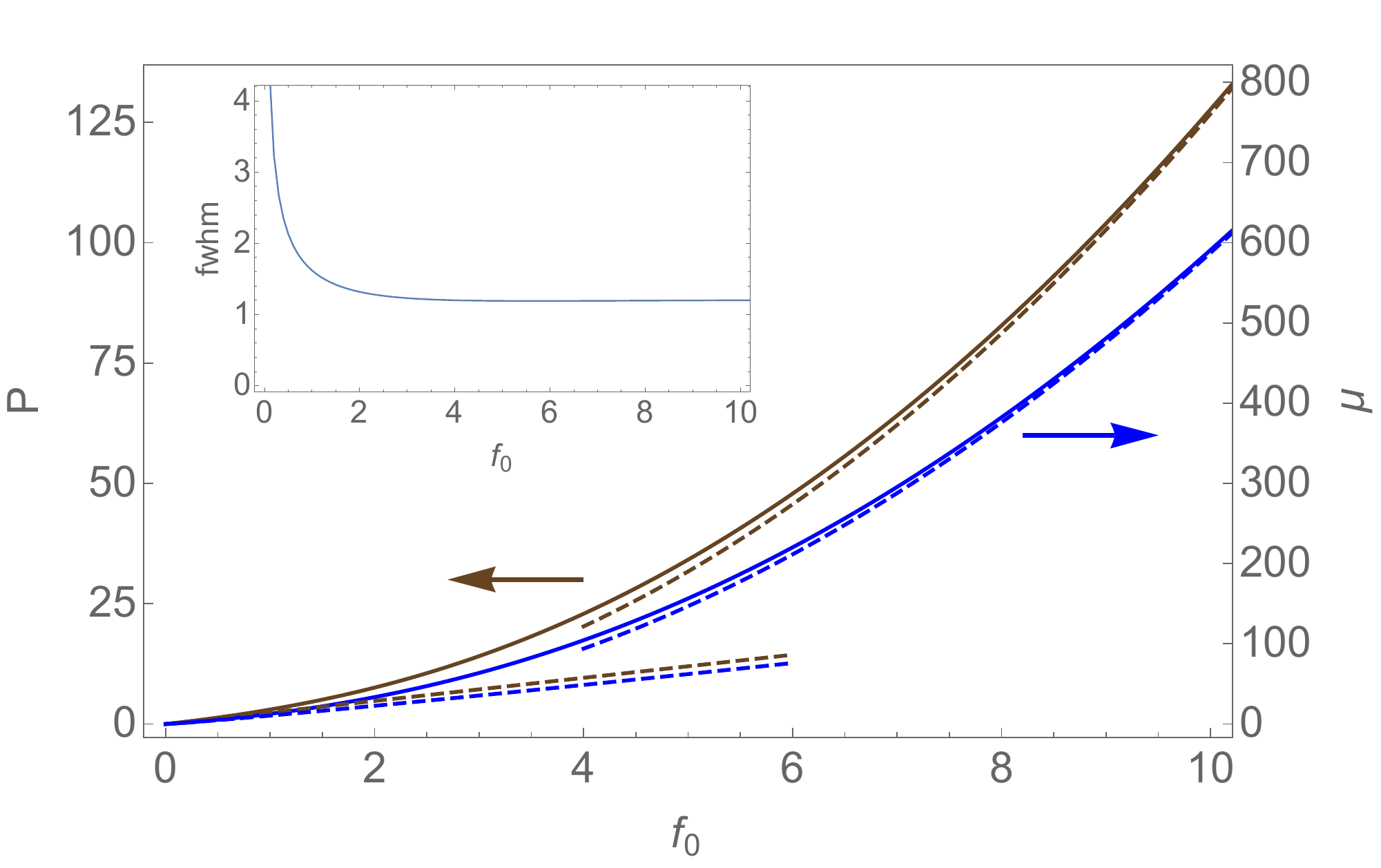}
\end{center}
\caption{[Color online] The adimensional power $P$ and propagation constant $\mu$ as a function of $f_0$ 
for the Schr\"odinger-Poisson equation with defocusing Kerr term $\lambda_K=-1$. Dashed lines
represent the asymptotic behaviors described in the text. In the inset, the full width at half maximum 
of the soliton solutions as a function of $f_0$.}
\label{fig6}
\end{figure}

Figure \ref{fig6} represents the power and propagation constant as
a function of $f_0$
for
 the family of ground state solutions with $\lambda_K=-1$.
 For small $f_0$, we have $P\approx 2.40 f_0$, $\mu\approx 10.53 f_0 + 1.20 f_0 \log f_0$
 as in the focusing case, since the Kerr term is unimportant in this limit.
 For large $f_0$, both nonlinearities play a 
decisive role. We have found by directly fitting the numerical data that 
 both $P$ and $\mu$ grow quadratically $P \approx 1.27 f_0^2$,
 $\mu \approx 5.88 f_0^2$. A remarkable feature
 is that the size of the soliton solutions tends to a constant for large $f_0$.
More precisely, the
 full width at half maximum of the distribution, defined as
 $f(r={\rm fwhm}/2)= f_0 / \sqrt2$ asymptotes to 
 $\lim_{f_0 \to \infty} {\rm fwhm} \approx 1.21$, see the inset of 
 figure \ref{fig6}.

Regarding oscillations mediated by boundary conditions, the dynamics with $\lambda_K=-1$
is rather similar to the one with focusing Kerr term because this effect is linked to the nonlocal nonlinearity
while the local nonlinear term only affects
 the shape of the soliton itself but is hardly related to its overall motion.

\section{IV. Soliton interactions and dark matter analogues}

In this section, we provide several examples of the dynamics of interacting solitons
 by numerically analyzing Eqs. (\ref{parax2}), (\ref{poisson2}).
These simulations are relevant for the propagation of light in thermo-optical media,
as described in section II. Moreover, they can be considered as analogues of
galactic
dark matter dynamics. In the context of the scalar field dark matter ($\psi$DM), soliton interactions
have been studied in different situations, including head-on collisions
 \cite{guzman1,guzman2,Paredes:2015wga,PhysRevD.93.103535,cotner},
dipole-like structures \cite{PhysRevD.94.043513}
and soliton mergers \cite{schiveprl,PhysRevD.94.043513}. These works deal with 
Eqs. (\ref{parax2}), (\ref{poisson2})  with one more transverse dimension $d=3$, but, as we 
will show, there are many qualitative similarities with the $d=2$ case.
We remark that the dynamics of soliton collisions is of great importance in cosmology,
since the wave-like evolution of $\psi$DM provides different outcomes from those of 
particle-like dark matter scenarios. Thus, it may furnish a way of 
improving our understanding of the nature and dynamics of dark matter, allowing us to
discriminate between different scenarios and to make progress in one of the most
important open problems of fundamental physics. For instance,
wave interference at a galactic scale can induce large offsets between dark matter distributions
and stars that might explain some recent puzzling observations \cite{Paredes:2015wga}.

Some
technical details of the numerical methods were briefly 
explained for Fig. \ref{fig3} above.
In order to generate  initial conditions, we 
 use the $f(r)$ profiles of the
eigenstates discussed in section III:
\begin{equation}
\psi|_{z=0} = \sum_{i=1}^{n_{sol}} f_i( |{{\bf x} - {\bf x}_i|}) e^{i ({\bf v}_i \cdot {\bf x} + \phi_i)}
\label{init_cond}
\end{equation}
where the sum runs over a number $n_{sol}$
 of initial solitons with initial positions 
${\bf x}_i$, phases $\phi_i$ and ``velocities'' ${\bf v}_i $.
From now on, for $dx_s/dz$ we use the word velocity , which is appropriate for
matter waves. In the optical setup, this quantity is of course the angle of propagation with
respect to the axis.
Boldface characters represent two-dimensional vectors ${\bf x}=(x,y)$, etc.
In the examples displayed below, boundary conditions $\Phi=0$ are set at the perimeter of a square of 
side $L=20$ and center at ${\bf x}=0$.

\subsection{a. Head-on collisions}

We start by analyzing the encounter of two solitons of the same power
 with equal phases. In the cosmological three-dimensional setup, this kind
of problem has been addressed in 
\cite{guzman1,guzman2,Paredes:2015wga,PhysRevD.93.103535}.
Qualitative results are very similar in the present $d=2$ optical setup.
What happens during evolution largely depends on the initial relative
velocity as we describe below.

\begin{figure}[h!]
\begin{center}
\includegraphics[width=\columnwidth]{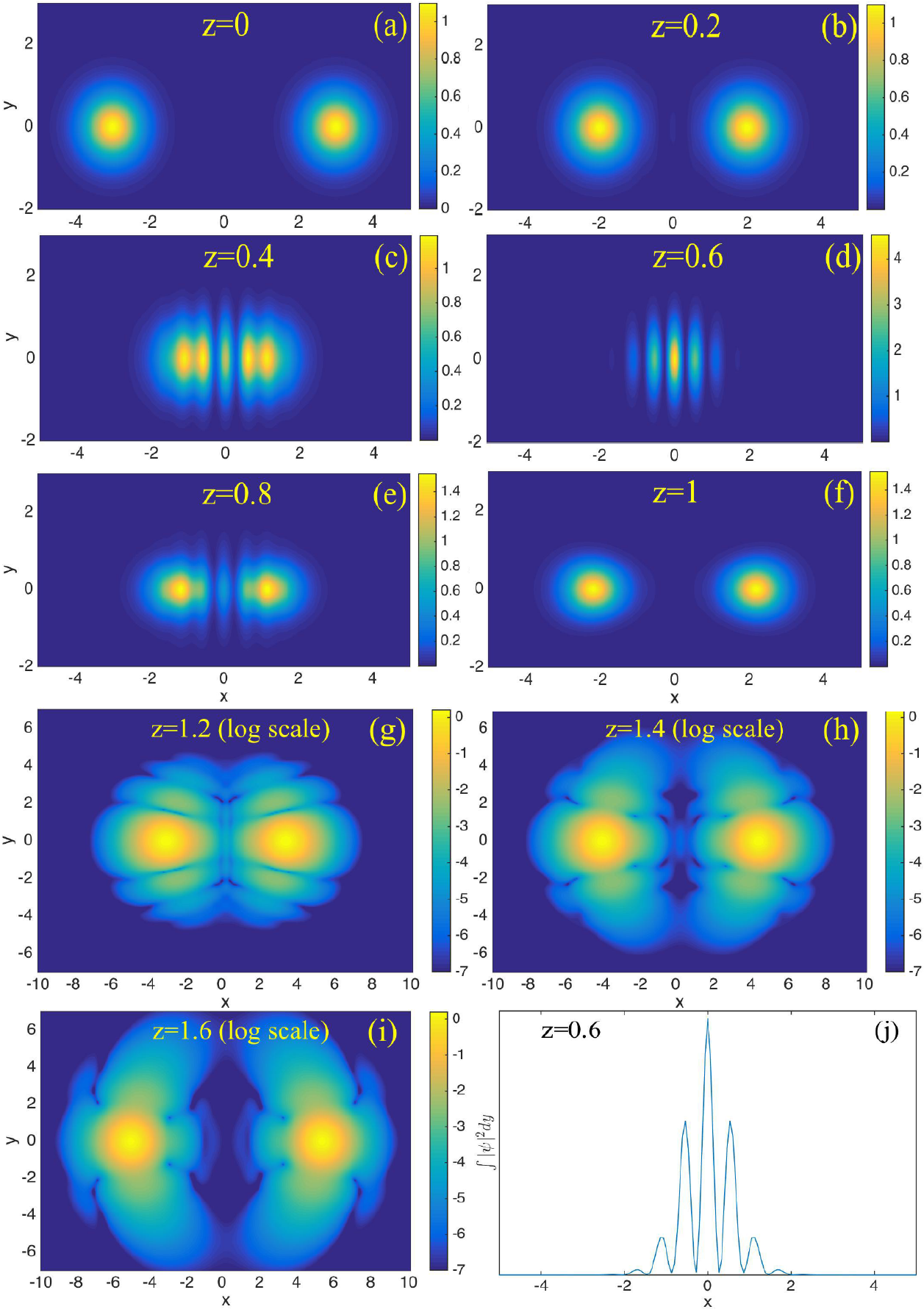}
\end{center}
\caption{[Color online] Two solitons crossing each other and producing an interference pattern when they meet. The images (a)-(f) are contour plots of $|\psi|^2(x,y)$ at different values of the propagation
distance $z$ for a 
numerical simulation of (\ref{parax2}), (\ref{poisson2}) with
$\lambda_K=1$.
Panels (g)-(i)  are contour plots of $\log_{10}(|\psi|^2(x,y))$ for the subsequent evolution.
We use logarithmic scale in order to visualize the energy flowing away from the soliton
centers (a meager $1\%$ in the example).
Panel (h) depicts the interference pattern at $z=0.6$, integrated in $y$.
Initial conditions are set by Eq. (\ref{init_cond})
with $n_{sol}=2$, $\phi_i=0$, $-x_1=x_2=3$, $v_1=-v_2=5$, $f_{0,1}=f_{0,2}=1.049$ ($P_1=P_2\approx 2$).
}
\label{fig7}
\end{figure}

For large velocities, where for large we mean that ${|\bf v}|$ is larger than the inverse size of the
initial structures, the solitons cross each other. During the collision, 
a typical interference fringe pattern is produced. Moreover, some tiny fraction of
energy is radiated away from the
solitons.
We stress that we are using the word soliton in a loose sense, since the theory is
not integrable and therefore even if the solitary waves cross each other, they do not come out
undistorted. This behavior is depicted in figure \ref{fig7}. After the crossing, the non-local
attraction and the self-interaction mediated by boundary conditions pull the solitons against
each other again and, depending on the particular case, this might result in a
second collision.

\begin{figure}[h!]
\begin{center}
\includegraphics[width=\columnwidth]{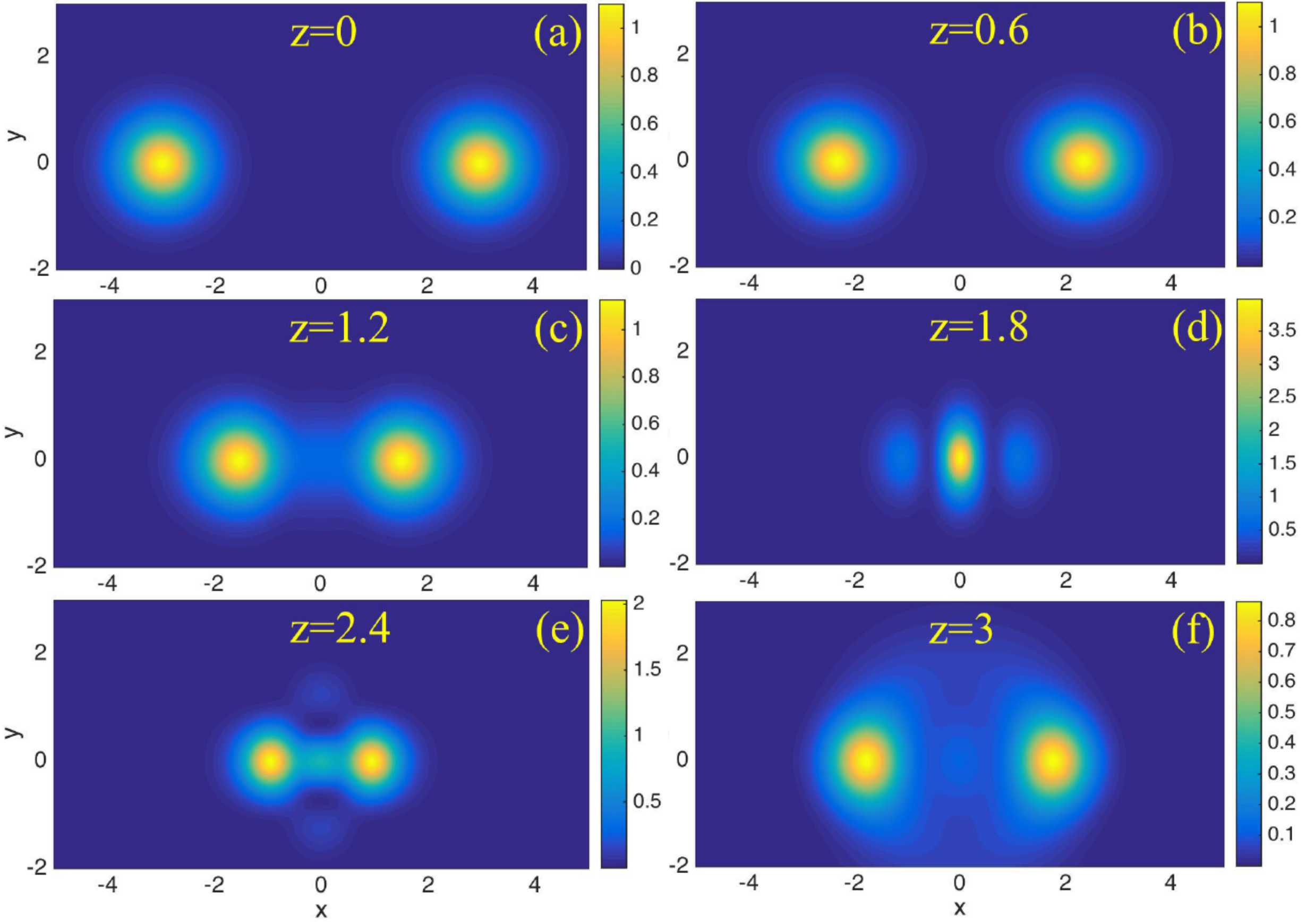}
\end{center}
\caption{[Color online] Two solitons crossing each other with
insufficient velocity to produce several interference fringes. 
All parameters are as in Fig. \ref{fig7} except for $v_1=-v_2=1$.}
\label{fig8}
\end{figure}

For intermediate velocities, solitons also cross each other but the associated wavelength
is too small to generate a pattern with multiple fringes. Figure \ref{fig8} shows an example.

For small velocities, the solitons merge in a  fashion similar to subsection c below.

\subsection{b. Dipole-like configuration}

As it is well known in different contexts, solitons in phase opposition repel each other. 
Possible consequences of this fact for galactic clusters
were explored in \cite{Paredes:2015wga}. 
In order to illustrate the fact, we consider a dipole-like structure: two solitons of the same size and
power in phase opposition bounce back from each. Due to the nonlocal nonlinearity, they attract each
other until they bounce back again and so on (see \cite{PhysRevD.94.043513} for similar
consideration in the dark matter context). 
There is a competition between the attraction due to 
the Poisson term and the repulsion due to  wave destructive interference. The Kerr term
contributes to attraction or repulsion depending on its sign. 

At this point, it is important to comment on the boundary conditions for the 
wave, since for large propagations part of the electromagnetic energy can reach the
bound of the domain. We will consider absorbing conditions, which are the best suited for cosmological
analogues. They can be implemented introducing at the borders a material with
 an imaginary
part of the refractive index, but with the same real part as the one of the bulk.
Mathematically, it can be modeled by introducing a ``sponge'', as discussed for the
three-dimensional dark matter case in \cite{guzman1,0004-637X-645-2-814,PRD69,PRD74}.
This amounts to adding a term:
\begin{eqnarray}
-\frac{i}{4}V_0 \left(4 - \tanh\frac{x+\gamma}{\delta} + \tanh\frac{x-\gamma}{\delta}+\right.
\nonumber\\
\left.
+
 \tanh\frac{y+\gamma}{\delta} + \tanh\frac{y-\gamma}{\delta}
\right) \psi
\label{sponge}
\end{eqnarray}
to the right hand side of (\ref{parax2}). This is a smooth version of a step function 
\cite{PRD69}. $\gamma$ fixes the position of the step and $\delta$ controls how steep the step is.
 In our simulations we fix $\gamma=0.4$, $\delta=0.2$,
$V_0=1$.

\begin{figure}[h!]
\begin{center}
\includegraphics[width=\columnwidth]{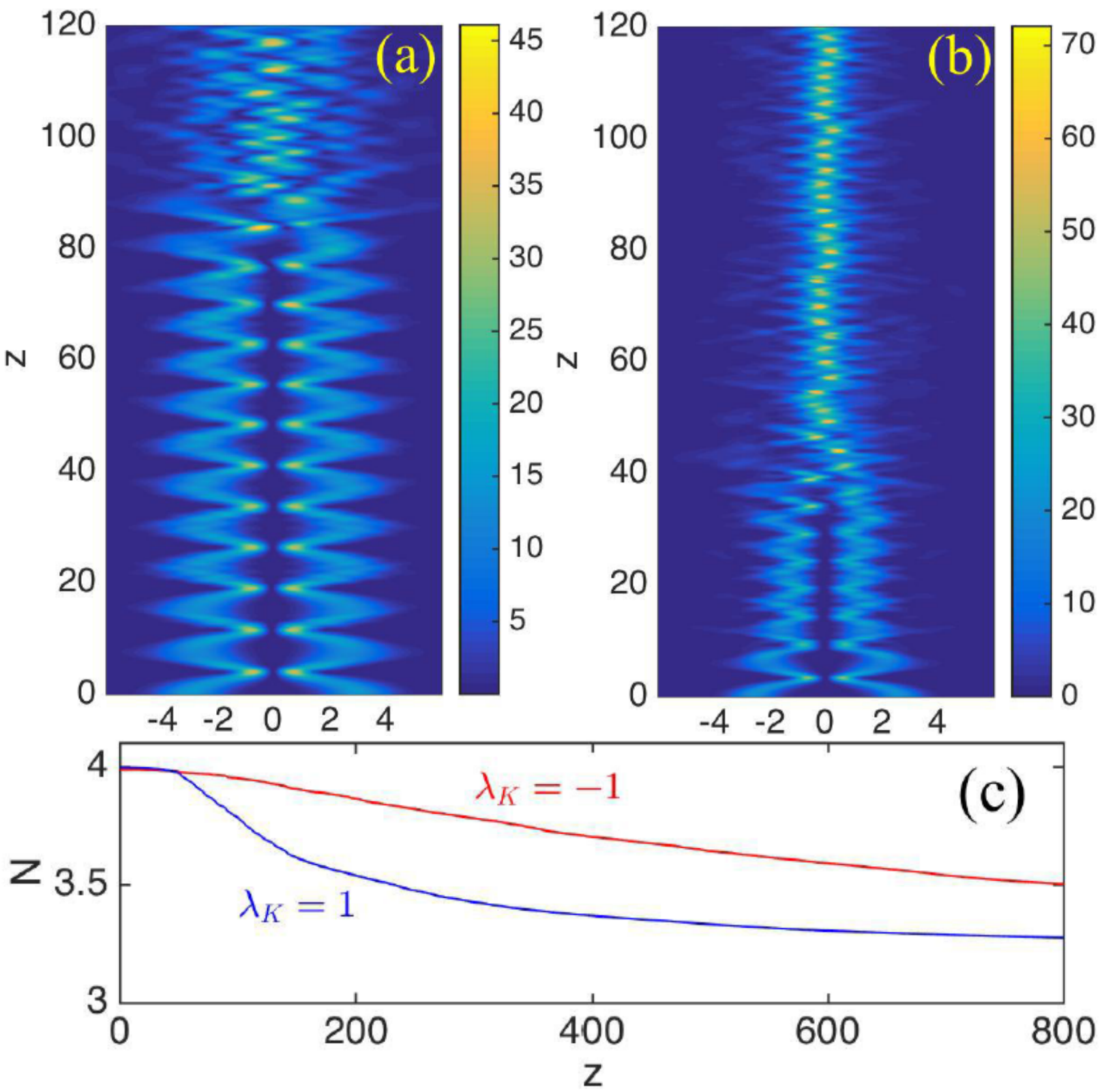}
\end{center}
\caption{[Color online] Solitons in phase opposition bouncing against each other in a dipole-like structure
and finally getting destabilized. We show contour plots of $|\psi|^2(x,z)$ 
integrated along $y$ for 
 simulations of (\ref{parax2}), (\ref{poisson2}) 
with $n_{sol}=2$,
$v_1=v_2=0$, $\phi_2-\phi_1=\pi$. 
In panel (a), $\lambda_K=-1$ (defocusing Kerr term), 
$-x_1=x_2=3$, 
$f_{0,1}=f_{0,2}=0.7097$ ($P_1=P_2=2$).
In panel (b), $\lambda_K=1$ (focusing Kerr term),
$-x_1=x_2=2.6$, $f_{0,1}=f_{0,2}=1.409$ ($P_1=P_2=2$).
Panel (c) portrays how the norm decreases due to the 
``sponge potential'', showing that when the dipolar structure breaks down and
the soliton merge, a fraction of energy is radiated away.
}
\label{fig9}
\end{figure}

Figure \ref{fig9} shows an example of the bounces of a dipolar configuration, comparing the evolution for
cases with $\lambda_K= \pm 1$ and the same soliton power. It is shown, that, eventually, the 
bouncing pattern becomes unstable and the solitons merge. As one could expect, this happens before
for focusing Kerr nonlinearity $\lambda_K=+1$. Notice, however, that the values of $z$ reached
in fig. \ref{fig9} are much larger than those of the other figures of this section. This means that 
the instabilities only become manifest for long propagations.
In fig \ref{fig9}, we also plot the $z$-evolution of the norm $N=\int |\psi|^2 dxdy$, which decreases
when the radiation approached the boundary because of the absorbing condition described
above. This is the analogue of having scalar radiation flowing away from the region of interest in
$\psi$DM. The figure shows that it starts happening when the instability breaks the initial solitons.
The system eventually evolves into a pseudo-stationary state, similar to the one described
in the next subsection.

We have not found any stable dipolar structure, but this aspect might deserve further
research.

\subsection{c. Soliton mergers}

In the $\psi$DM model of cosmology, the  outcome of the merging of solitons has
important consequences for the galactic dark matter distributions. In 
\cite{schiveprl,PhysRevD.94.043513}, it was proven that the final configuration
of such a process is a new, more massive, soliton (to be identified with a galactic core)
surrounded by an incoherent distribution of matter with its density decreasing with a power law.
The gravitational attraction prevents that this halo is radiated away.
Here, we will show that a very similar behavior takes place in $d=2$, paving the way for 
optical experiments partially mimicking galactic mergers.

\begin{figure}[h!]
\begin{center}
\includegraphics[width=\columnwidth]{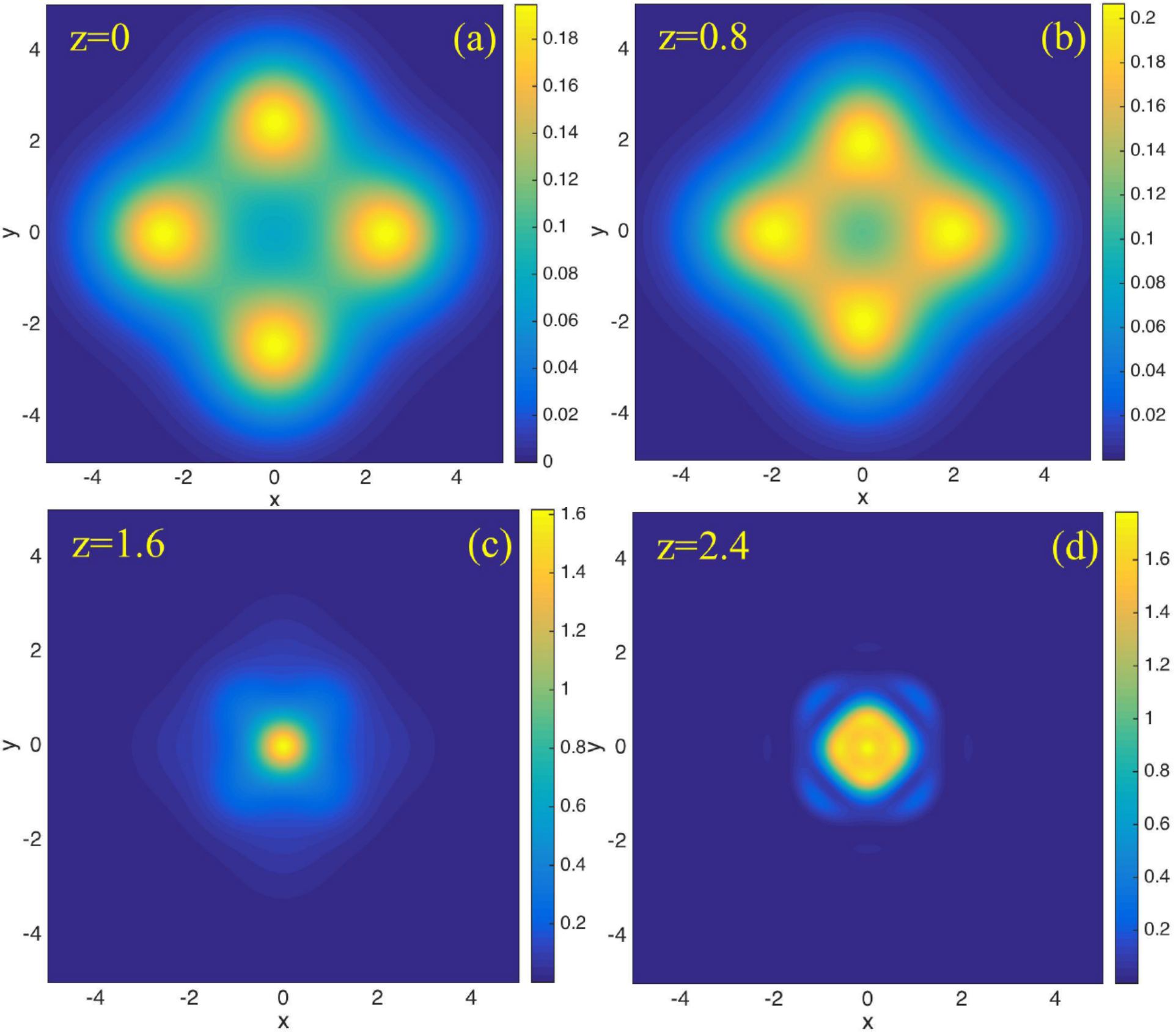}
\end{center}
\caption{[Color online] 
Four merging solitons. The images are contour plots of $|\psi|^2(x,y)$ at 
different values of $z$ for a
 simulation  with
$\lambda_K=1$.
Initial conditions are set by Eq. (\ref{init_cond})
with $n_{sol}=4$, $\phi_i=0$, $-x_1=x_2=-y_3=y_4=2.6$, $v_i=0$,
 $f_{0,i}=0.4$ (the power of the entire configuration is $P\approx 5.14$).}
\label{fig10}
\end{figure}

Figure \ref{fig10} depicts an example of the initial stages of evolution of four equal merging solitons.
The Kerr nonlinearity has been taken to be focusing and the total power to be below Townes'
threshold in order to avoid a possible collapse.
The solitons rapidly coalesce and form a peaked narrow 
structure with a faint distribution of power around it.

\begin{figure}[h!]
\begin{center}
\includegraphics[width=\columnwidth]{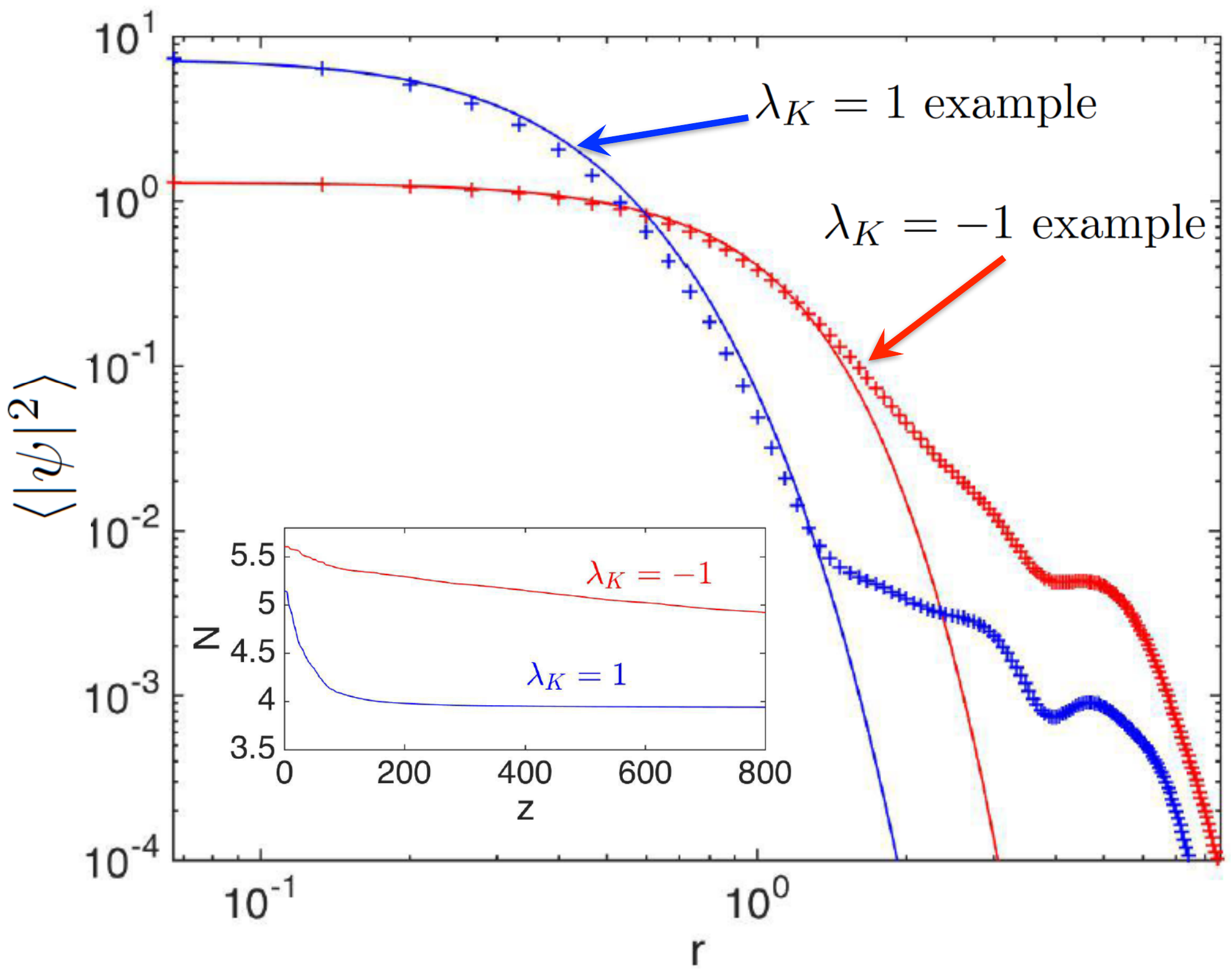}
\end{center}
\caption{[Color online] Doubly logarithmic plot
depicting the asymptotic power distribution for the simulation of
figure \ref{fig10} ($\lambda_K=1$) and another simulation with 
$\lambda_K=-1$ (four initial solitons with $f_0=4$, $\phi_i=0$,
 $-x_1=x_2=-y_3=y_4=3$, $P\approx 5.61$). 
The crosses are the result of the numerical computations and the
solid lines represent the soliton profiles (section III) 
with the same energy density at $r=0$.
In the vertical axis, $\langle|\psi|^2\rangle$ 
represents an
average in $z$, taken over the interval $z\in [600,800]$ and along the
line $y=0$, $x>0$, but the result  depends only mildly in this particular choices.
In the inset, we depict the $z$-evolution of the norm $N=\int |\psi|^2 dx dy$
due the absorbing potential (\ref{sponge}), showing
that some energy is radiated away during the merging and then, slowly, the configuration
tends to a pseudo-stationary situation.
}
\label{fig11}
\end{figure}

Continuing the numerical evolution of Eqs. (\ref{parax2}), (\ref{poisson2}) to large values of $z$,
a pseudo-stationary situation is attained, with an oscillation 
about a soliton profile at its center and incoherent
radiation around it. This is shown in
figure \ref{fig11}, where we depict an average in $z$ of the density profile depending on
the distance to the center. We do the computation for the absorbing boundary condition
presented in section IV.b.
We also include an example with defocusing Kerr term $\lambda_K=-1$.

The graphs show that, roughly speaking, the merging results in a soliton surrounded
by a halo of energy trapped by ``gravity''. 
Part of the initial energy has been radiated away during the merging of the solitonic structures.
Their qualitative agreement
with those of \cite{schiveprl,PhysRevD.94.043513} is apparent.
The
force mediated by boundary conditions for $\Phi$ affects the result but does not change the
qualitative picture with respect to the three-dimensional case.

\section{V. Conclusions}

We have analyzed the Schr\"odinger-Poisson equations (\ref{parax2}), 
(\ref{poisson2}) in two
dimensions ($d=2$) in the presence of a Kerr term. We have limited the discussion to
positive sign for the Poisson term.
The model is relevant to describe laser propagation in thermo-optical 
materials, among other physical systems.
With radially symmetric boundary conditions for $\Phi$,
we have found uniparametric families of radially symmetric stable solitons.
When the Kerr term is focusing, the family interpolates between the solution without 
Kerr term and the Townes profile. For defocusing Kerr term, the family interpolates between the solution without Kerr term and solitons which asymptotically tend to a particular finite size.
The fixed boundary conditions induce effective forces which push the solitons towards the
center of the material.

Regarding interactions, the Poisson term produces attraction at a distance, resembling gravity.
This fact has been studied experimentally \cite{segev}. 
Both the local and nonlocal nonlinearities shape the solitons and the
results of dynamical evolution. However, we remark that in soliton collisions, a prominent role
is played by the wave nature of Schr\"odinger equation. 
As in many nonlinear systems, interference
fringes appear for appropriate initial conditions and there is attraction/repulsion for 
phase coincidence/opposition.

We have remarked that the same equations, in one more dimension ($d=3$), are the basis of the
scalar field dark matter ($\psi$DM) model of cosmology, which relies on the hypothesis
of the existence of a cosmic Bose-Einstein condensate of an ultralight axion.
In this scenario, the physics of solitons is connected to phenomena taking place at length
scales comparable to galaxies. 

Certainly, there are differences between the cosmological $d=3$ case and the possible
laboratory $d=2$ setups. First of all, we have not considered situations with evolution
of the cosmic scale factor. Furthermore, the Poisson interaction is stronger in smaller 
dimension and the monopolar ``gravitational force'' decays as $1/r$ in $d=2$ rather
than as $1/r^2$.  It is of particular importance the role of boundary 
conditions. In the Universe, one typically has open boundary conditions.
In a laboratory experiment, they have to be specified at a finite distance from the
center. For the Poisson field (temperature), the simplest is to make it constant at the
borders of the sample, even if this generates restoring forces towards the center.
For the electromagnetic wave, the best is to introduce an absorbing element that prevents
reflections towards the center of the radiation reaching the borders. This simulates
the sponge typically used in the numerical $3d$ computations. In any case, the effects
related to boundary conditions are reduced by taking larger two-dimensional sections of the
optical material.
We see no obstruction, apart from cost, to utilizing large pieces of
glass in this kind of experiment.
Let us also remark that, apart from the analogy discussed here, boundary conditions
are a useful turning knob in optical experiments with nonlocal nonlinearities.

Despite these discrepancies and the disparate physical scales involved,
there are apparent strong similarities between the
families of self-trapped waves, their stability and their interactions in the
$d=2$ and $d=3$ cases. 
We emphasize that 
this resemblance holds with or without Kerr term, corresponding to the presence
or absence of non-negligible local self-interactions of the scalar in the cosmological setup.
We hope 
that
these considerations will pave the way for the experimental engineering of
optical experiments that introduce analogues of dark matter dynamics in the
$\psi$DM scenario.

\section{Appendix: Scaling the equation to its canonical form}

Consider the equation:
\begin{eqnarray}
i a_1 \frac{\partial \tilde \psi}{\partial \tilde z}&=&-\frac12 a_2 \tilde\nabla^2 \tilde\psi - 
\lambda_K
a_3 |\tilde\psi|^2\tilde\psi + a_4
\tilde\Phi \tilde\psi 
\label{parax3}\\
\tilde\nabla^2 \tilde\Phi &=& a_5 |\tilde\psi|^2
\label{poisson3}
\end{eqnarray}
where the $a_i>0$ are constants and tilded quantities correspond to dimensionful
coordinates, potential and wave function. Equations (\ref{parax3}), (\ref{poisson3})
are transformed into the canonical dimensionless form (\ref{parax2}), (\ref{poisson2}) by 
the following rescalings.
\begin{eqnarray}
&&\tilde z= \frac{C a_1 a_3}{a_2 a_4 a_5} z, \qquad\quad\
(\tilde x,\tilde y)= \left(\frac{C a_3}{a_4 a_5}\right)^\frac12(x,y) \nonumber\\
&&\tilde \psi= \left(\frac{a_2 a_4 a_5}{C a_3^2}\right)^\frac12 \psi, \qquad\
\tilde \Phi= \frac{a_2 a_5}{C a_3} \Phi .
\label{changegen}
\end{eqnarray}

\acknowledgments

{\it Acknowledgements.} We thank Alessandro Alberucci, Jisha Chandroth Pannian 
and Camilo Ruiz for
useful comments. We also thank two anonymous referees that helped us improving
the discussions on the physical implementation and the analogy to cosmological
setups.
This work is supported by grants FIS2014-58117-P and FIS2014-61984-EXP
from Ministerio de Econom\'\i a y Competitividad and grants GPC2015/019
and EM2013/002 from Xunta de Galicia.

\end{document}